\documentclass[twocolumn,astrosymb]{aastex631}

\usepackage{textcomp}
\usepackage{amssymb}
\usepackage{amsmath}
\usepackage{comment}
\usepackage{wasysym}
\usepackage{CJK}
\usepackage{verbatim}
\usepackage{enumerate}

\shorttitle{stellar populations in NGC 2992}
\shortauthors{Xu}

\begin{document}

\begin{CJK*}{UTF8}{gbsn}


\title{Spatially resolved analysis of Stellar Populations in NGC 2992: Impact of AGN feedback}

\correspondingauthor{Junfeng Wang}
\email{jfwang@xmu.edu.cn}

\author[0000-0003-0970-535X]{Xiaoyu Xu (许啸宇)}
\affiliation{School of Astronomy and Space Science, Nanjing University, Nanjing 210023, China}
\affiliation{Key Laboratory of Modern Astronomy and Astrophysics, Nanjing University, Nanjing 210023, China}
\affiliation{Department of Astronomy, Xiamen University, Xiamen, Fujian 361005, China}

\author[0000-0003-4874-0369]{Junfeng Wang}
\affiliation{Department of Astronomy, Xiamen University, Xiamen, Fujian 361005, China}

\author[0000-0003-0355-6437]{Zhiyuan Li}
\affiliation{School of Astronomy and Space Science, Nanjing University, Nanjing 210023, China}
\affiliation{Key Laboratory of Modern Astronomy and Astrophysics, Nanjing University, Nanjing 210023, China}

\author{Yanmei Chen}
\affiliation{School of Astronomy and Space Science, Nanjing University, Nanjing 210023, China}
\affiliation{Key Laboratory of Modern Astronomy and Astrophysics, Nanjing University, Nanjing 210023, China}

\begin{abstract}

In NGC 2992, a galaxy-scale ionized gas outflow driven by AGN has long been recognized, yet its impact on the host galaxy has remained elusive. 
In this paper, we utilize data from the archival Very Large Telescope (VLT)/MUSE to present a spatially resolved analysis of stellar populations in this galaxy.
Two different stellar population templates are employed to fit the stellar continuum, allowing us to determine the light-weighted stellar age, metallicity, the fraction of the young stellar population (age $<100$ Myr, $P_{\rm Y}$), and the average age and metallicity of $P_{\rm Y}$.
Our results reveal the presence of a very young stellar population ($\leq40$ Myr) within the dust lane and nearly along the galaxy's major axis.
The light-weighted stellar age and the fraction of $P_{\rm Y}$ show negative trends along the major and minor axes. 
The average age and metallicity of $P_{\rm Y}$ present positive trends with increasing distance, except along the northern direction of the major axis.
Within the circumnuclear region ($<1$ kpc), the distribution of the young stellar population is spatially anti-correlated with the AGN outflow cone. 
The highest fraction of $P_{\rm Y}$ is observed at the outskirts of the nuclear radio bubble in the northern region near the nucleus.
Considering the coupling efficiency and timescales, we propose that the AGN outflow in this galaxy may exert both negative and positive feedback on its host.
Additionally, the star formation and the AGN activities could be attributed to the interaction between NGC 2992 and NGC 2993.

\end{abstract}


\section{Introduction} \label{sec:intro}

It is widely anticipated that active galactic nuclei (AGNs) can exert a significant influence on the surrounding environment and the evolution of their host galaxies. 
This influence is manifested through the release of energy in the form of radiation and the generation of powerful outflows \citep[e.g.][]{2012ARA&A..50..455F}.
Observations have provided substantial evidence for the presence of AGN outflows in both neutral (atomic and molecular) and ionized phases \citep[e.g.,][]{2013ApJ...776...27V,2014MNRAS.441.3306H,2019MNRAS.483.4586F,2019A&A...622A.146M}.

In a negative feedback scenario, AGN outflows have the capability to heat and/or expel cold gas, leading to a depletion of gas supplies that subsequently suppresses star formation (SF) \citep[e.g.][]{1998A&A...331L...1S,2008ApJS..175..356H,2012RAA....12..917S}. 
Some studies have shown that AGN outflows exhibit a spatial anti-correlation with star formation regions within the host galaxies of some high-redshift AGNs, providing direct evidence of quasar-driven outflows quenching star formation in galaxies \citep[e.g.][]{2012A&A...537L...8C,2015ApJ...799...82C}.
However, these studies are based on the H$\alpha$ emission line, which may not be a good tracer of the obscured SF. 
Follow-up studies based on the far-infrared (FIR) luminosity did not confirm the anti-correlation \citep[][]{2020MNRAS.492.3194S,2021MNRAS.505.5469S}.
In many luminous AGNs, the molecular gas depletion time caused by AGN outflows is shorter compared to star-forming galaxies, indicating a significant impact of AGN outflows on both star formation and the gas reservoir within the host galaxy \citep[e.g.][]{2017A&A...601A.143F,2019MNRAS.483.4586F}.

Alternatively, some works have suggested that AGN outflows can also trigger or enhance SF processes, known as positive feedback
\citep[e.g.][]{2013ApJ...772..112S,2017MNRAS.468.4956Z}.
In some luminous AGNs, regions of active star formation can be found surrounding the AGN outflows, indicating that the outflows compress the gas and trigger star formation at the edge of the outflow \citep[e.g.][]{2015ApJ...799...82C,2023A&A...678A.127V}.
Using near-infrared adaptive optics integral field spectrograph (IFS) data, \citet{2007ApJ...671.1388D} found some evidence of a young stellar population (10-300 Myr) in the central region ($<1$ kpc) of several nearby Seyfert galaxies.
In some of these Seyfert galaxies (e.g. Mkr 231, Circinus, NGC 7469), AGN-driven outflows in the central region have been detected \citep[e.g.][]{2010A&A...518L.155F,2019A&A...622A.146M,2021ApJ...906L...6R,2023MNRAS.519.5324K}.
\citet{2009MNRAS.392.1295L} identified a very young stellar population ($<15$ Myr) in the nuclear region ($\sim200$ pc) of Mrk 231, where a powerful cold gas outflow was observed \citep{2010A&A...518L.155F,2015A&A...583A..99F}.
These observational findings might suggest the presence of an AGN positive feedback scenario.
Recently, \citet{2023A&A...678A.127V} discovered a young stellar population ($<$100--150 Myr) co-located with the ionized gas loop in the Teacup galaxy, with the star formation timescale aligning with the expansion time of the loop, indicating a positive AGN feedback in this galaxy.

\begin{figure}[ht!]
\includegraphics[width=0.48\textwidth]{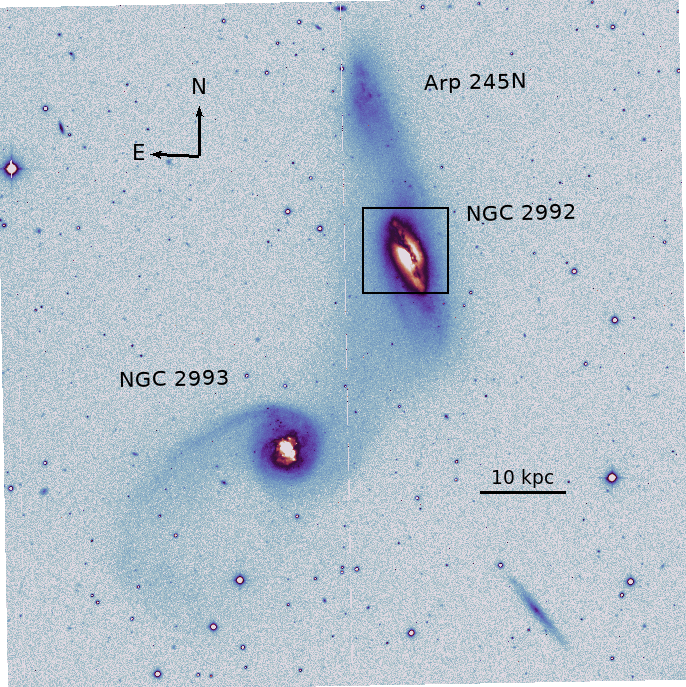}
\caption{An optical (V band) image of the interacting system Arp 245 (from NED).
The black box denotes the FoV of the VLT/MUSE data.
}
\label{fig:arp_245}
\end{figure}

NGC 2992 is a member of an interacting system Arp 245.
Figure~\ref{fig:arp_245} shows a V band image of the interacting system Arp 245 observed using the du Pont telescope (from NED\footnote{https://ned.ipac.caltech.edu}), which includes NGC 2992, a star-forming galaxy NGC 2993, and a tidal dwarf galaxy Arp 245N. 
Notably, there is a prominent dust lane along the major axis of NGC 2992 and a diffuse bridge connecting NGC 2992 and NGC 2993 \citep[e.g.,][]{2000AJ....120.1238D}. 
According to \citet{2000AJ....120.1238D}, this system is in an early stage of interaction, approximately 100 Myr after perigalacticon.

NGC 2992 is a nearly edge-on spiral galaxy situated at $z=0.00771$ \citep{1996ApJS..106...27K} which corresponds to a distance of 32.5 Mpc ($1\arcsec \sim150$ pc). 
Within this galaxy resides an AGN renowned for its remarkable variability.
Notably, the 2-10 keV flux of NGC 2992 exhibits fluctuations of a factor of $\sim 30$ over periods of weeks to months \citep[e.g.][]{2018MNRAS.478.5638M,2022MNRAS.514.2974M}. 
Additionally, the optical classification of this AGN varies between Seyfert 1.5 and 1.9 \citep{2008AJ....135.2048T}.

NGC 2992 features two prominent figure-eight-shaped radio bubbles that extend to the northwest and southeast in the circumnuclear region ($<1$ kpc).  
These radio bubbles, approximately $8\arcsec$ in size, exhibit an axis misalignment of about $26^{\circ}$ from the galaxy's major axis \citep{1984ApJ...285..439U,2023A&A...679A..88Z}.  
The figure-eight-shaped radio bubbles are likely expanding bubbles driven by AGN as suggested by \citet{2001A&A...378..787G} and \citet{2005ApJ...628..113C}.
Recently, \citet{2022ApJ...938..127X} suggested that the majority of the circumnuclear extended soft X-ray emission ($r \sim 750 \rm\, pc$) originates from AGN-driven outflows. 
On the galactic scale, \citet{2017MNRAS.464.1333I} reported C-band (5--7 GHz) observations of NGC 2992, presenting compelling evidence of a bipolar radio outflow detected in linearly polarized emission. 

In the optical band, the detection of a biconical AGN outflow aligned with the minor axis of the NGC 2992 galactic disk has been reported in various studies \citep[e.g.][]{2001AJ....121..198V,2019A&A...622A.146M,2020AJ....159..167L,2022MNRAS.511.2105K, 2023A&A...679A..88Z}. 
This outflow extends over approximately $7\rm\, kpc$ \citep[][]{2023A&A...679A..88Z}. 
Based on the spatially resolved Baldwin, Phillips, and Telervich \citep[BPT,][]{1981PASP...93....5B,1987ApJS...63..295V} diagnostic diagrams, the primary ionization source of the outflow was confirmed to be the AGN \citep[][]{2019A&A...622A.146M,2020AJ....159..167L,2021MNRAS.502.3618G,2022MNRAS.511.2105K}.

Some evidence of interaction between AGN outflows and the ISM had been found in NGC 2992 \citep[e.g.][]{2017MNRAS.464.1333I,2022ApJ...938..127X,2023A&A...679A..88Z}, suggesting that AGN outflows may impact star formation in the host galaxy. 
The ongoing interaction between NGC 2992 and NGC 2993 could lead to gas inflow and trigger star formation in these galaxies. 
However, due to AGN ionization contamination, classical emission-line star formation rate (SFR) probes (e.g., H$\alpha$ emission) are not suitable for use in active galaxies. 
Utilizing stellar population synthesis through full spectral fitting can provide insights into the young star formation in AGN host galaxies. 
Comparing the distribution of young stellar populations with AGN outflows in NGC 2992 could aid in exploring the relationship between AGN outflows and star formation. 
Furthermore, the distribution of young stellar populations may also indicate the impact of the galaxy interaction.

In this study, we present a spatially resolved analysis of NGC 2992 using the deep VLT/MUSE data.
Although this archival data has been analyzed in previous studies \citep[e.g.,][]{2019A&A...622A.146M,2020AJ....159..167L,2022MNRAS.511.2105K,2023A&A...679A..88Z}, our analysis focuses on the spatially resolved properties of the stellar populations and a potential connection between the AGN outflows and the stellar populations.
In Sec.~\ref{sec:data}, we describe the observations and reduction of the data. 
The spatially resolved properties of stellar populations and the radial profiles are shown in Sec.~\ref{sec:result}.
In Sec.~\ref{sec:discussion}, discussions about the impacts of the AGN and galaxy merger are presented. 
Lastly, a summary is provided in Sec.~\ref{sec:summary}.

\section{Observations and Data Reduction} \label{sec:data}

\begin{deluxetable*}{ccc}
\tablenum{1}
\tablecaption{SSPs used in the stellar population synthesis. \label{tab:ssp}}
\tablewidth{0pt}
\tablehead{
\colhead{Model} & BC03 & FSPS
}
\startdata
Evolution track & Padova (1994) & MIST \\
\hline
Spectral library & STELIB & MILES \\
\hline
Metallicity ($\rm Z_{\odot}$)  & 0.02, 0.2, 0.4, 1, 2.5 & 0.018, 0.18, 0.56, 1, 1.8 \\
\hline
 & 0.001, 0.003, 0.005, 0.01, 0.03, 0.055 & 0.001, 0.003, 0.005, 0.01, 0.03, 0.06  \\
Age (Gyr) & 0.064, 0.081, 0.1, 0.29, 0.57, 0.90 & 0.08, 0.1, 0.31, 0.63, 0.79  \\
 & 3.0, 5.0, 11, 15, 20 & 3.2, 5.0, 12.6, 15.8 \\
\enddata
\end{deluxetable*}

\begin{figure*}[ht!]
\includegraphics[width=\textwidth,trim=0 20 0 20]{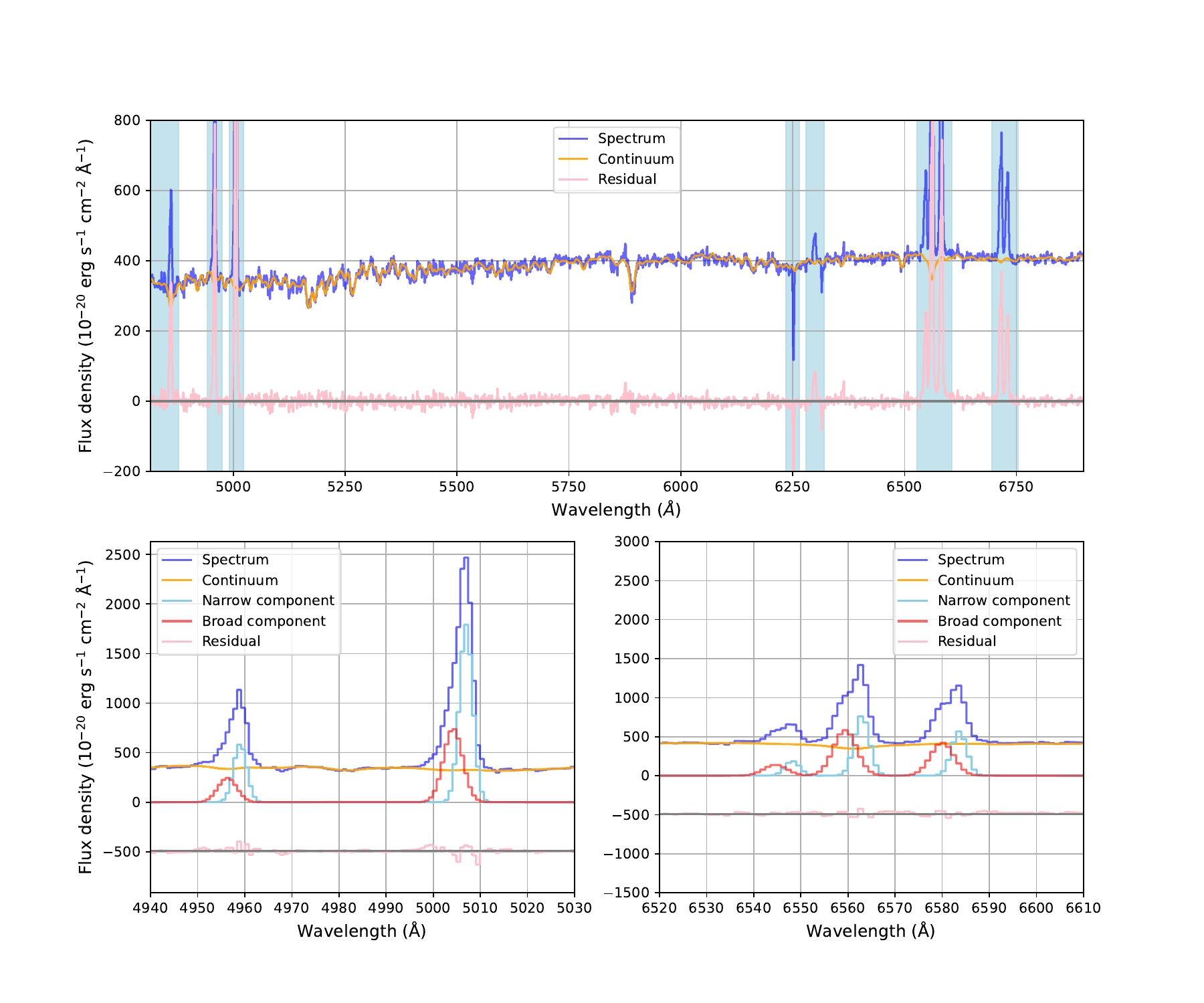}
\caption{Top panel: An example of spectra (shown in blue) with the continuum (orange line) from a spaxel at $\Delta$RA $=\rm -4.\arcsec0$ and $\Delta$Dec $=\rm 2.\arcsec6$ from the nucleus. 
The shaded regions show the masked bands when fitting the continuum.
The lower left panel shows the spectrum zoomed in on [O~{\sc{iii}}] emission lines, and the lower right panel H$\alpha$ and [N~{\sc{ii}}]. 
Sky-blue lines and red lines denote narrow and broad emission lines, respectively.
For better visualization, the residual (shown in pink) is arbitrarily shifted below zero.
}
\label{fig:spec_example}
\end{figure*}

\subsection{MUSE data}\label{subsec:muse}

NGC 2992 was observed in the wide field mode (WFM) by MUSE \citep[][]{2010SPIE.7735E..08B,2014Msngr.157...13B} in 2015, with a total exposure time of $\sim 10\rm\, ks$, under program 094.B-0321(A) (PI: Marconi). 
A field of view (FoV) of $\sim 1$ arcmin$^{2}$ with spatial sampling of $\rm 0.\arcsec2$ pixel$^{-1}$ is provided, while the spatial resolution of these data was limited by seeing ($\sim 1\arcsec$). 
The wavelength coverage is $\rm 4650-9300\, \AA$ with a mean resolution of R $\sim$ 3000.
We downloaded data from ESO archive\footnote{$\rm http://archive.eso.org/eso/eso\_archive\_main.html$} and used EsoReflex \citep{2013A&A...559A..96F} with MUSE workflow v2.8.5 \citep{2020A&A...641A..28W} to reduce the data.
The final data cube was corrected for Milky Way extinction, using $A_{\rm V} = 0.166$ \citep[][]{2011ApJ...737..103S} and the extinction law of \cite{1999PASP..111...63F}.

\subsection{Stellar populations synthesis}\label{subsec:ste-pop syn}

We employed the Voronoi binning method \citep[][]{2003MNRAS.342..345C} to spatially rebin the data cube, aiming to achieve a mean signal-to-noise ratio (S/N) threshold of 40 over the continuum emission in the wavelength range of $\rm 4800-7000\,\AA$. 
The Penalized Pixel-Fitting method \citep[PPXF;][]{2023MNRAS.526.3273C} was used to fit the spectrum of every Voronoi bin within the same wavelength range.
The stellar population models used in this study include the \citet[hereafter BC03]{2003MNRAS.344.1000B} model and Flexible Stellar Population Synthesis \citep[FSPS;][]{2009ApJ...699..486C,2010ApJ...712..833C} model.
BC03 model use the Padova evolution track \citep[][]{1993A&AS..100..647B} and the STELIB spectral library \citep{2003A&A...402..433L}, while FSPS model employs the MESA Isochrones \& Stellar Tracks \citep[MIST][]{2016ApJS..222....8D,2016ApJ...823..102C,2019ApJS..243...10P} and the MILES spectral libraby \citep{2006MNRAS.371..703S} (Table~\ref{tab:ssp}).
Both models adopted the same \cite{2003PASP..115..763C} initial mass function (IMF).
We employed a set of 85 representative single stellar populations (SSPs) with 17 ages ranging from 1 Myr to 20 Gyr and 5 metallicities ranging from 0.02 $\rm Z_{\odot}$ to 2.5 $\rm Z_{\odot}$ provided from BC03 model. 
For FSPS templates, we selected a set of SSPs with age and metallicity coverage as similar as possible to the SSPs from the BC03 model.
Increasing the number of SSPs did not yield a significant improvement in the results.
The complete list of SSPs is shown in Table~\ref{tab:ssp}.
Even though the Voronoi binning method was used, the stellar continuum fitting is bad at the edge of FoV due to contamination from sky lines.
Hence, we exclude the binned spaxels with a continuum level below $4\times10^{-19}\rm\ erg\ cm^{-2}\ s^{-1} \AA^{-1}$ at $5100\rm\ \AA$ from the results of stellar population synthesis.

We obtained the stellar kinematics and properties of the stellar populations by masking the gas emission lines (with a masking width of 1000 km/s) and fitting the spectrum in each bin.
The extinction curve from \citet{2000ApJ...533..682C} was used to fit the stellar reddening.
To obtain uncertainties of the stellar population parameters, we employed the bootstrapping method \citep[e.g.][]{DAVIDSON2008162}. 
Firstly, the spectrum was fitted using PPXF.
Secondly, a new spectrum was generated by randomly adding $\pm$residual to the best-fit model at each wavelength channel.
This new spectrum was then fitted using PPXF.
The bootstrapping process was repeated 20 times, and one standard deviation of every resulting stellar population parameter (e.g. age, metallicity, light weights) from these 20 sets was used as the error.

For convenience, we used two simplified stellar populations with light-weighted age ranges: $t \leq 100\,\rm Myr$ ($P_{\rm Y}$) and $100\,{\rm Myr} \leq t$ ($P_{\rm O}$).
In this work, we will focus on the young stellar population ($P_{\rm Y}$), because the timescale of this population is comparable to the dynamical timescale of the ionized outflow (see Sec.~\ref{subsec:timescale}).
$f_{\rm Y}$ denotes the fraction of $P_{\rm Y}$, calculated by summing the light weights of $P_{\rm Y}$ within each spaxel.


\subsection{Emission lines}\label{subsec:emission lines}

Previous studies have examined the optical emission lines in NGC 2992 using the identical MUSE data \citep[][]{2019A&A...622A.146M, 2020AJ....159..167L, 2022MNRAS.511.2105K, 2023A&A...679A..88Z}. 

After subtracting the stellar continuum, two Gaussian components were used to fit the main gas emission lines: H$\beta$, [O~{\sc{iii}}]$\lambda\lambda$4959,5007, H$\alpha$, [N~{\sc{ii}}]$\lambda\lambda$6548,6583, and [S~{\sc{ii}}]$\lambda\lambda$6716,6731.
For the same component, the velocity $v$ and the velocity dispersion $\sigma$ of all emission lines share the same value.
The $\sigma$ range of the first component was limited to 50-200 km/s, while the $\sigma$ of the second component was set to be at least 50 km/s larger than that of the first component. 
Only components with S/N$\ge3$ are retained.
Figure~\ref{fig:spec_example} illustrates an example of spectral fitting extracted from a spaxel in the outflow cone (the red solid circle in Figure~\ref{fig:fsps_stepop}(a)).
The morphology of the first and the second components are very similar to the disk and the outflow component defined by \cite{2019A&A...622A.146M} and \cite{2022MNRAS.511.2105K}.
Hence, the first and the second components are referred to as the disk and the outflow component, respectively.

\section{Results} \label{sec:result}

\subsection{Stellar extinction and kinematics} \label{subsec:extinction & kinematics}

\begin{figure*}[ht!]
\includegraphics[width=\textwidth,trim=0 70 0 70]{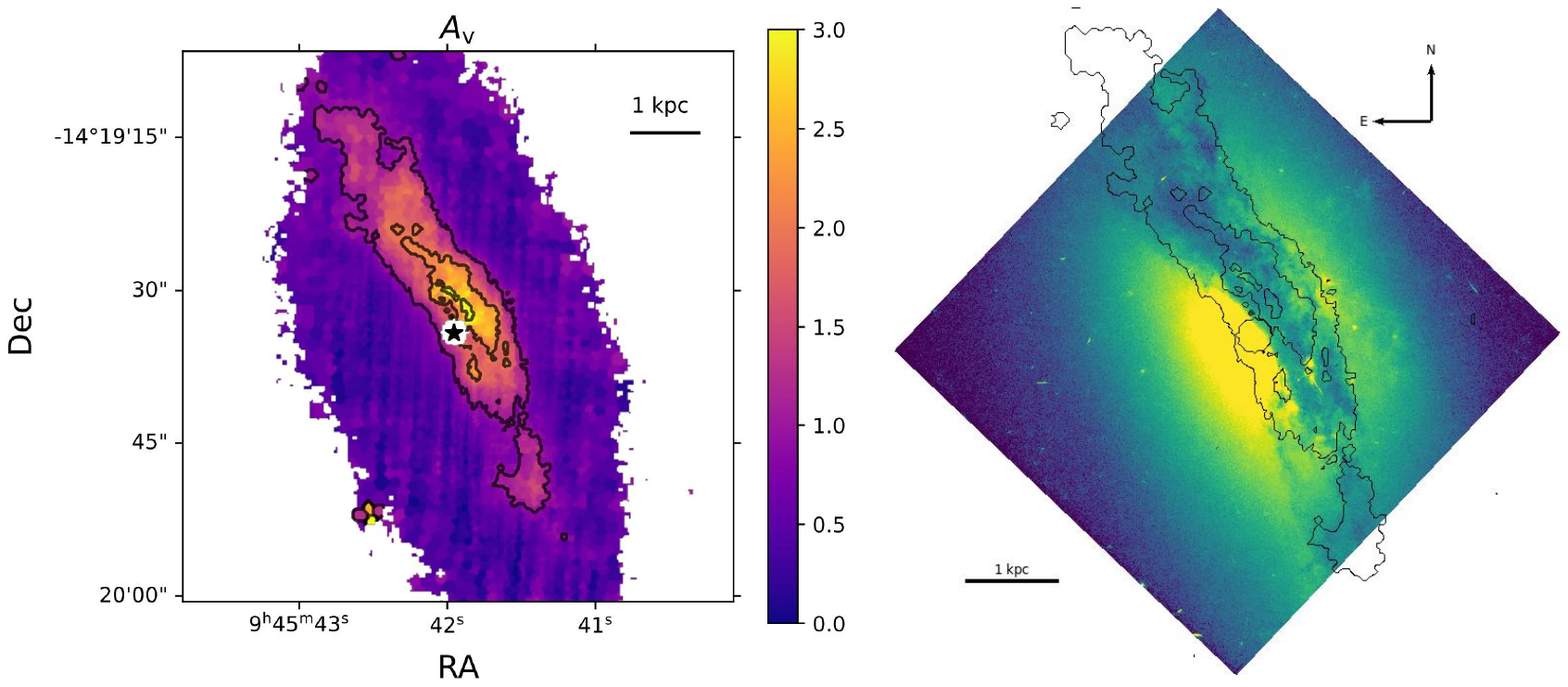}
\caption{Left: The extinction in V band $A_{\rm V}$ derived from stellar continuum fitting using FSPS templates.
Black contours represent $A_{\rm V} = 1.0,\ 2.0,\ 3.0$, respectively.
The nuclear region (with a radius $r<1.2 \arcsec$) is masked.
The black star denotes the peak of [O~{\sc{iii}}] disk component.
Binned spaxels with stellar continuum below $4\times10^{-19}\rm\ erg\ cm^{-2}\ s^{-1} \AA^{-1}$ at $5100\rm\ \AA$ are excluded.
Right: HST/WFPC2 V-band image using the F606W filter \citep{1998ApJS..117...25M} superposed with the contours of $A_{\rm V}$ in the left panel.
}
\label{fig:extinction}
\end{figure*}

\begin{figure*}[ht!]
\includegraphics[width=\textwidth]{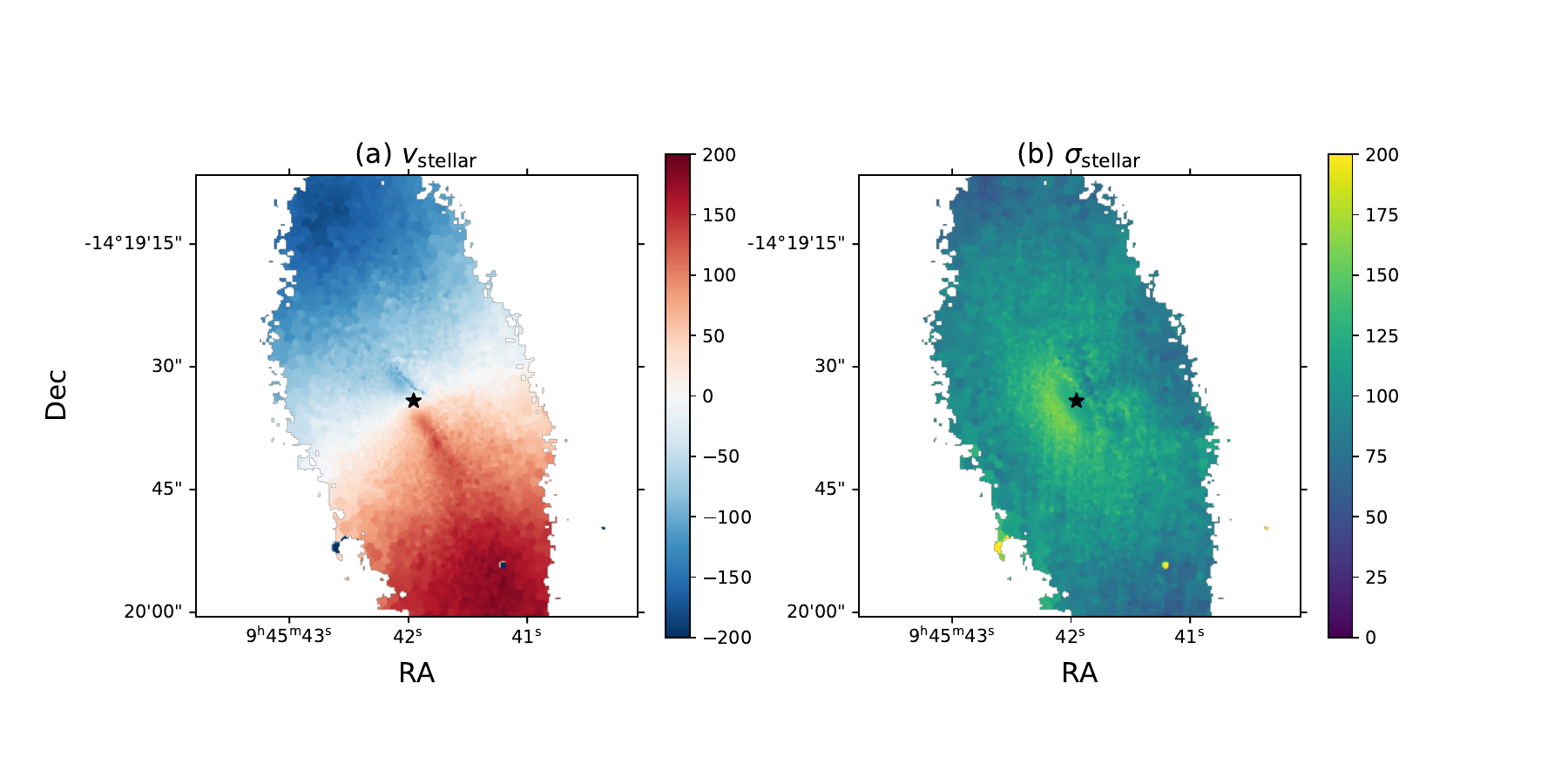}
\caption{Stellar kinematics map derived from stellar continuum fitting using FSPS templates.
The black star denotes the peak of [O~{\sc{iii}}] disk component.
Binned spaxels with stellar continuum below $4\times10^{-19}\rm\ erg\ cm^{-2}\ s^{-1} \AA^{-1}$ at $5100\rm\ \AA$ are excluded.
}
\label{fig:kinematics}
\end{figure*}

In the Hubble Space Telescope (HST) V-band image (right panel of Figure~\ref{fig:extinction}) from \cite{1998ApJS..117...25M}, a prominent dust lane is observed nearly along the major axis of NGC 2992.
The extinction in V band $A_{\rm V}$ obtained from the stellar continuum fitting aligns spatially with the morphology of this dust lane (left panel of Figure~\ref{fig:extinction}).
\cite{2019A&A...622A.146M} and \cite{2023A&A...679A..88Z} also calculated $A_{\rm V}$ from the Balmer decrement ($\rm H\alpha/H\beta$), which is higher than the $A_{\rm V}$ obtained in this study.
Additionally, \cite{2023A&A...679A..88Z} identified a molecular gas disk with an inclination angle of 80 degrees using ALMA CO(2-1) data, which coincides with the major axis of NGC 2992. 
The dust lane also exhibits rough consistency with the molecular disk discovered by \cite{2023A&A...679A..88Z}.

Figure~\ref{fig:kinematics} displays the stellar kinematics derived from continuum fitting, revealing a rotational stellar disk.
For stars located at a distance of $2 \rm\,kpc$ from the nucleus, the rotation timescale $t_{\rm rot}$ is estimated to be $\sim 60\rm\,Myr$. 
This suggests that the distribution of stellar populations in the stellar disk, except for very young stars, will be smoothed or disrupted due to rotation.
The $A_{\rm V}$ and stellar kinematics maps derived from BC03 synthesis are consistent with the Figure~\ref{fig:extinction} and Figure~\ref{fig:kinematics}.

\subsection{Spatially resolved properties of stellar populations} \label{subsec:ste-pop map}

\begin{figure*}[ht!]
\includegraphics[width=1\textwidth,trim=0 50 30 50]{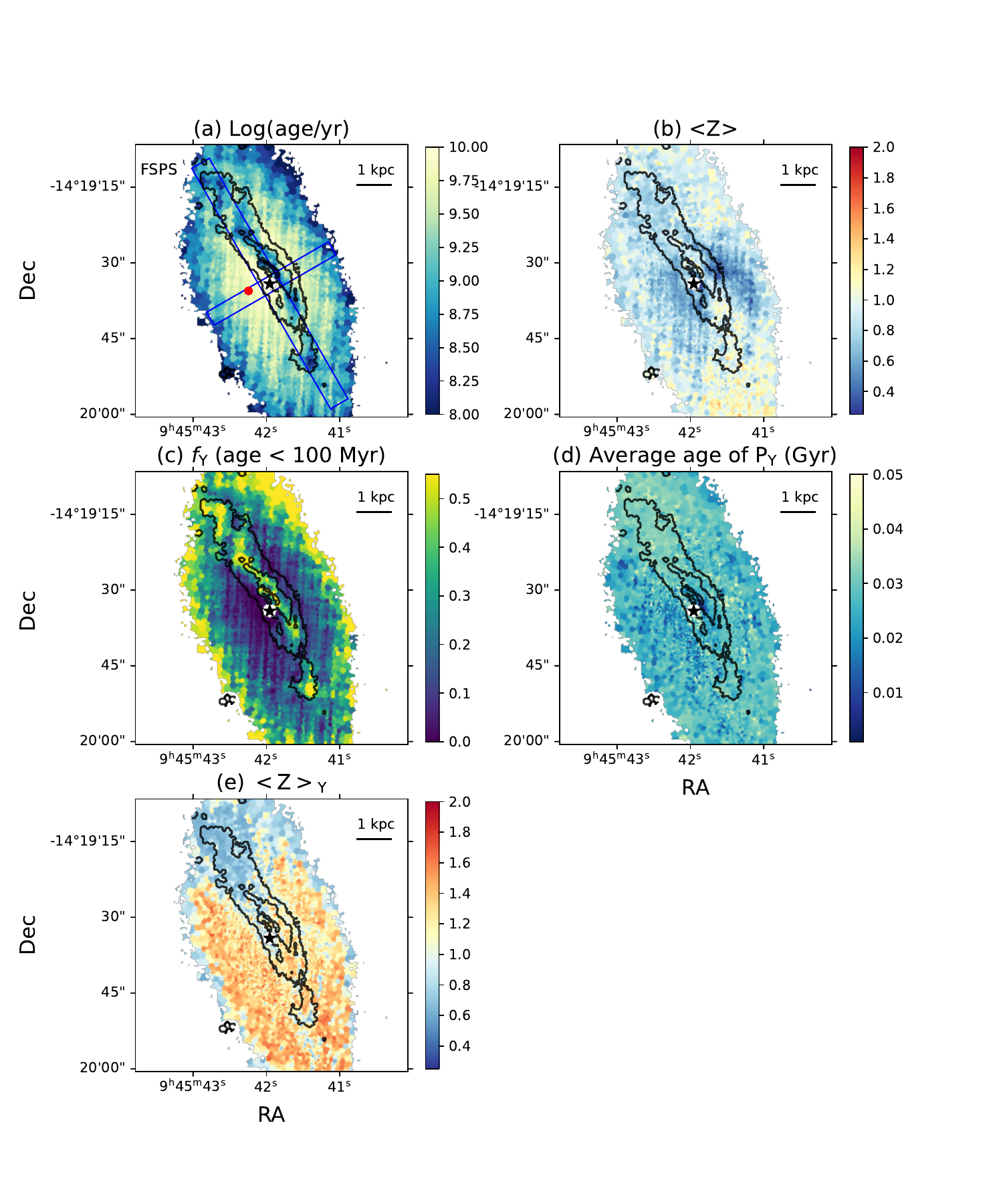}
\caption{Properties of stellar populations obtained from FSPS synthesis. 
(a): Light-weighted mean logarithmic stellar age. 
Two overlaid blue regions are used to derive radial profiles along the major and minor axes.
The red solid circle represents the location of the spectrum in Figure~\ref{fig:spec_example}.
(b): Light-weighted mean metallicity in a unit of $Z_{\odot}$.
(c): Fraction map of $P_{\rm Y}$ (stellar populations with age $\rm < 100\,Myr$).
(d): Light-weighted mean age of $P_{\rm Y}$.
(e): Light-weighted mean metallicity of $P_{\rm Y}$.
All maps have black contours representing the dust extinction ($A_{\rm V} = 1.0,\ 2.0,\ 3.0$).
The nuclear region (radius $r<1.2 \arcsec$) is masked.
The black star denotes the peak of [O~{\sc{iii}}] disk component.
}
\label{fig:fsps_stepop}
\end{figure*}

\begin{figure*}[ht!]
\includegraphics[width=1\textwidth,trim=0 50 30 50]{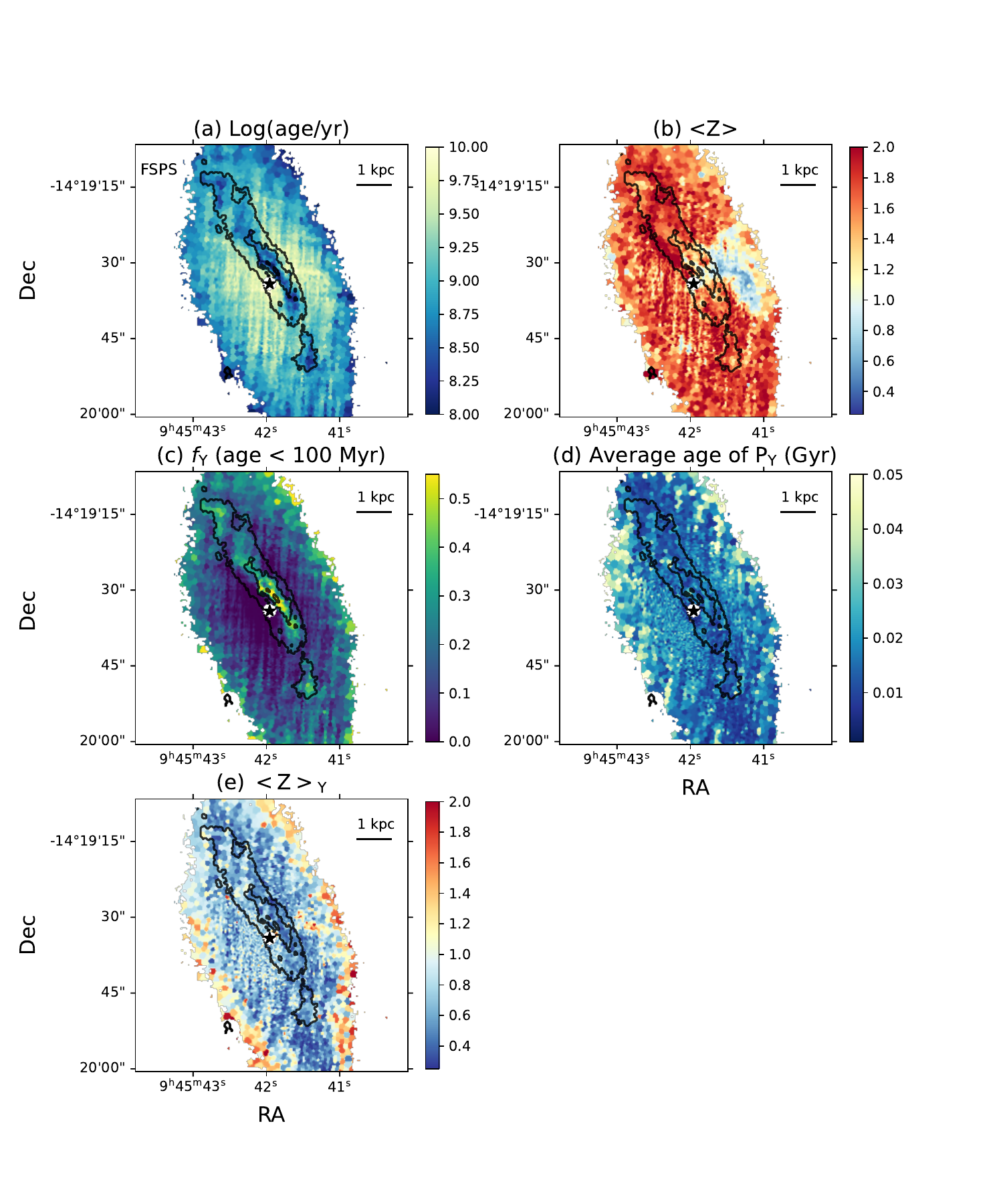}
\caption{Same as Figure~\ref{fig:fsps_stepop} but derived from the BC03 synthesis.
}
\label{fig:bc03_stepop}
\end{figure*}


In Figure~\ref{fig:fsps_stepop}, maps of the light-weighted average stellar age, light-weighted mean metallicity $<Z>$, the fraction of young stellar population $f_{\rm Y}$, and the mean age and metallicity ($<Z>_{\rm Y}$) of the young stellar population derived from the FSPS synthesis are shown. 
In the average stellar age map (Figure~\ref{fig:fsps_stepop}(a)), the stellar ages are lower within the dust lane.
Along the minor axis, the average age decreases as the distance from the nucleus increases.
In Figure~\ref{fig:fsps_stepop}(b), $<Z>$ derived from FSPS synthesis shows an asymmetry pattern: sub-solar $<Z>$ in the northern, eastern, and western regions, while solar $<Z>$ in the southern region.
Compared to Figure~\ref{fig:fsps_stepop}(c) and (e), the sub-solar $<Z>$ in the northern region is attributed to the very young stellar population.
In the eastern and western regions along the minor axis, the sub-solar $<Z>$ is mainly contributed by the old stellar populations.
In the $f_{\rm Y}$ map (Figure~\ref{fig:fsps_stepop}(c)), high $f_{\rm Y}$ is presented along the dust lane and at the edge of the stellar disk.
Some spectra in the edge of the disk have been checked and the continuum fitting is reliable.
However, $f_{\rm Y}$ is very low in the vicinity of the nuclear region, except for the west region, despite the presence of dust and substantial molecular gas \citep[][]{2023A&A...679A..88Z}.
Figure~\ref{fig:fsps_stepop}(d) shows that in most regions of the stellar disk, the young stellar population is very young ($< 40\rm\ Myr$).
A super-solar metallicity of this young stellar population is presented across the stellar disk except for the northern region (Figure~\ref{fig:fsps_stepop}(e)). 

The spatially resolved properties of stellar populations obtained from BC03 synthesis are presented in Figure~\ref{fig:bc03_stepop}.
The map of the mean stellar age (Figure~\ref{fig:bc03_stepop}(a)) is similar to the result obtained from FSPS synthesis (Figure~\ref{fig:fsps_stepop}(a)).
The $<Z>$ map (Figure~\ref{fig:bc03_stepop}(b)) shows super-solar metallicity throughout the galaxy, except for the western region.
In the $f_{\rm Y}$ map (Figure~\ref{fig:bc03_stepop}(c)), relatively high $f_{\rm Y}$ is presented along the dust lane and at the edge of the stellar disk.
Figure~\ref{fig:bc03_stepop}(d) and (e) show that in most regions of the stellar disk, the young stellar population is very young ($< 40\rm\ Myr$) and metal-poor.

Compared to the result from FSPS synthesis, $f_{\rm Y}$ obtained from BC03 synthesis is 0.1--0.2 lower in the dust lane and the edge of the stellar disk.
Additionally, the age of $P_{\rm Y}$ derived from BC03 synthesis is younger than that obtained from FSPS synthesis in the dust lane.
The $<Z>$ map obtained from FSPS synthesis (Figure~\ref{fig:fsps_stepop}(b)) reveals sub-solar metallicity throughout the galaxy, except for the southern region.
However, $<Z>$ derived from BC03 synthesis is different.
In Figure~\ref{fig:bc03_stepop}(b), only in the western region of NGC 2992 shows solar or sub-solar $<Z>$, likely originating from older stellar populations compared to Figure~\ref{fig:bc03_stepop}(e).
However, considering the galactic rotation, the distribution of the old stellar populations should be smoothed, casting doubt on the reliability of the pattern depicted in Figure~\ref{fig:bc03_stepop}(b).
The $<Z>_{\rm Y}$ maps derived from FSPS (Figure~\ref{fig:fsps_stepop}(e)) and BC03 (Figure~\ref{fig:bc03_stepop}(e)) synthesis exhibit significant differences.
The former shows super-solar $<Z>_{\rm Y}$ across the stellar disk except for the northern region, while the latter presents sub-solar $<Z>_{\rm Y}$ throughout the disk.
The disparity in metallicity derived from BC03 and FSPS stellar templates suggests that the metallicity measurements may not be reliable.

The consistent findings about the distribution and the mean age of $P_{\rm Y}$ from both FSPS and BC03 stellar templates indicate the presence of a very young stellar population ($<40$ Myr) within the dust lane.
\cite{2021MNRAS.502.3618G} analyzed the stellar populations in the nuclear region ($\leq3\arcsec$) in NGC 2992 using Gemini Multi-Object Spectrograph (GMOS) data.
They found a young ($<$100 Myr) stellar population with a relatively high light-weighted fraction ($10\%\leq f_{\rm Y}\leq30\%$) in the nuclear region.
Although the innermost region ($<1.2\arcsec$) is excluded in our work, the findings of \cite{2021MNRAS.502.3618G} are broadly consistent with our result.
\citet{2010A&A...519A..79F} also found evidence for a short burst of star formation (40-−50 Myr ago) in the nuclear region ($\sim3\arcsec \times 3\arcsec$) using data from the near-infrared integral field spectrograph SINFONI on the VLT.
Furthermore, \citet{2019A&A...622A.146M} presented the BPT diagrams of NGC 2992 based on the same MUSE data.
Their [S~{\sc{ii}}] BPT diagram reveals some star formation regions within the dust lane, which is also consistent with our results.

\subsection{Radial profiles along the major and minor axes} \label{subsec:radial profile}

\begin{figure*}[ht!]
\includegraphics[width=1\textwidth]{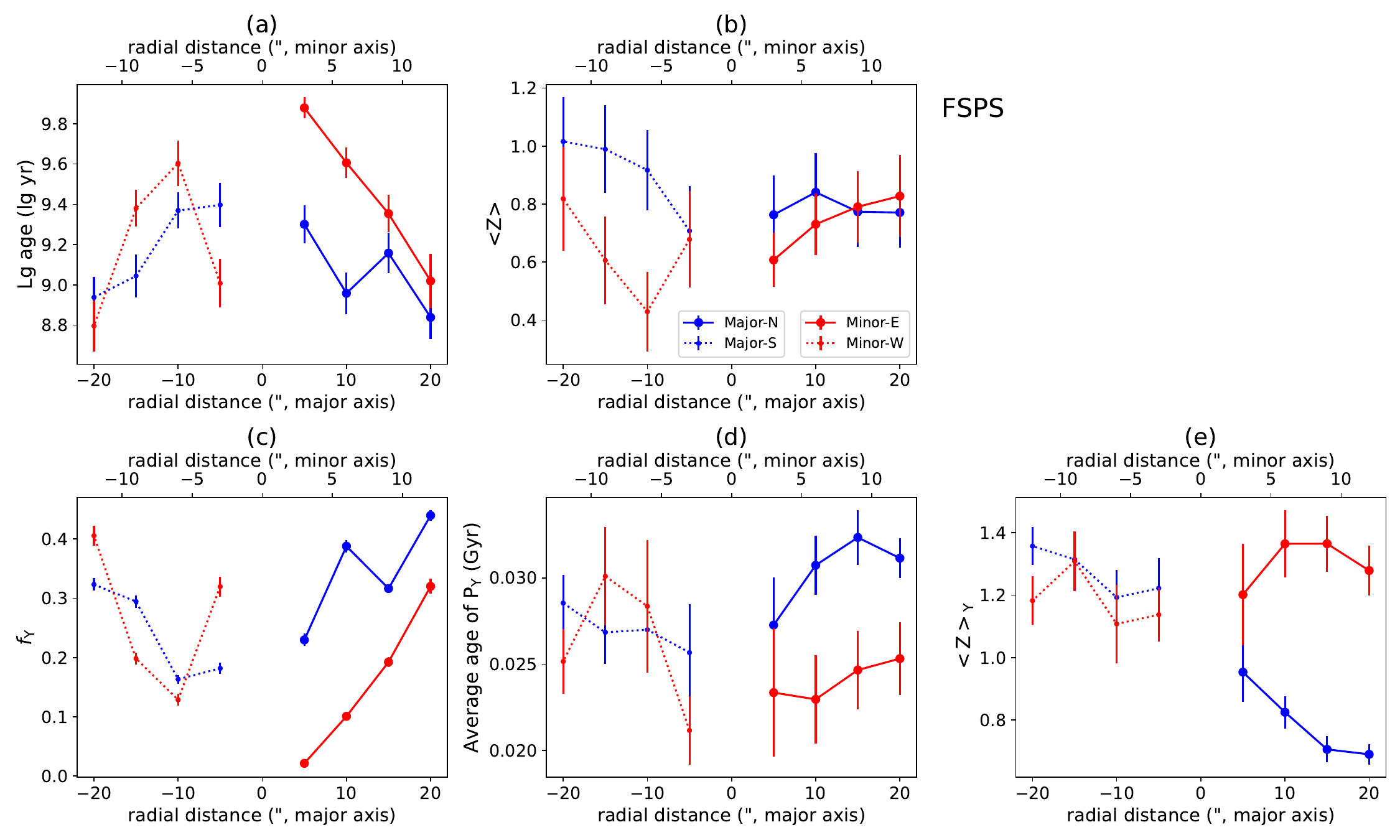}
\caption{Radial profiles of results derived from FSPS synthesis.
(a): Light-weighted mean logarithmic stellar age.
(b): Light-weighted mean metallicity in a unit of $Z_{\odot}$.
(c): $f_{\rm Y}$.
(d): Light-weighted mean age of $P_{\rm Y}$.
(e): Light-weighted mean age of $P_{\rm Y}$.
Major-N and Major-S refer to the radial profiles along the major axis towards the north and south directions, respectively. 
Similarly, Minor-E and Minor-W indicate the radial profile along the minor axis towards the east and west directions, respectively. 
Note that the radial distance along the major and the minor axes are presented in the bottom and top horizontal axes, respectively.
}
\label{fig:fsps_rprofile}
\end{figure*}

\begin{figure*}[ht!]
\includegraphics[width=1\textwidth]{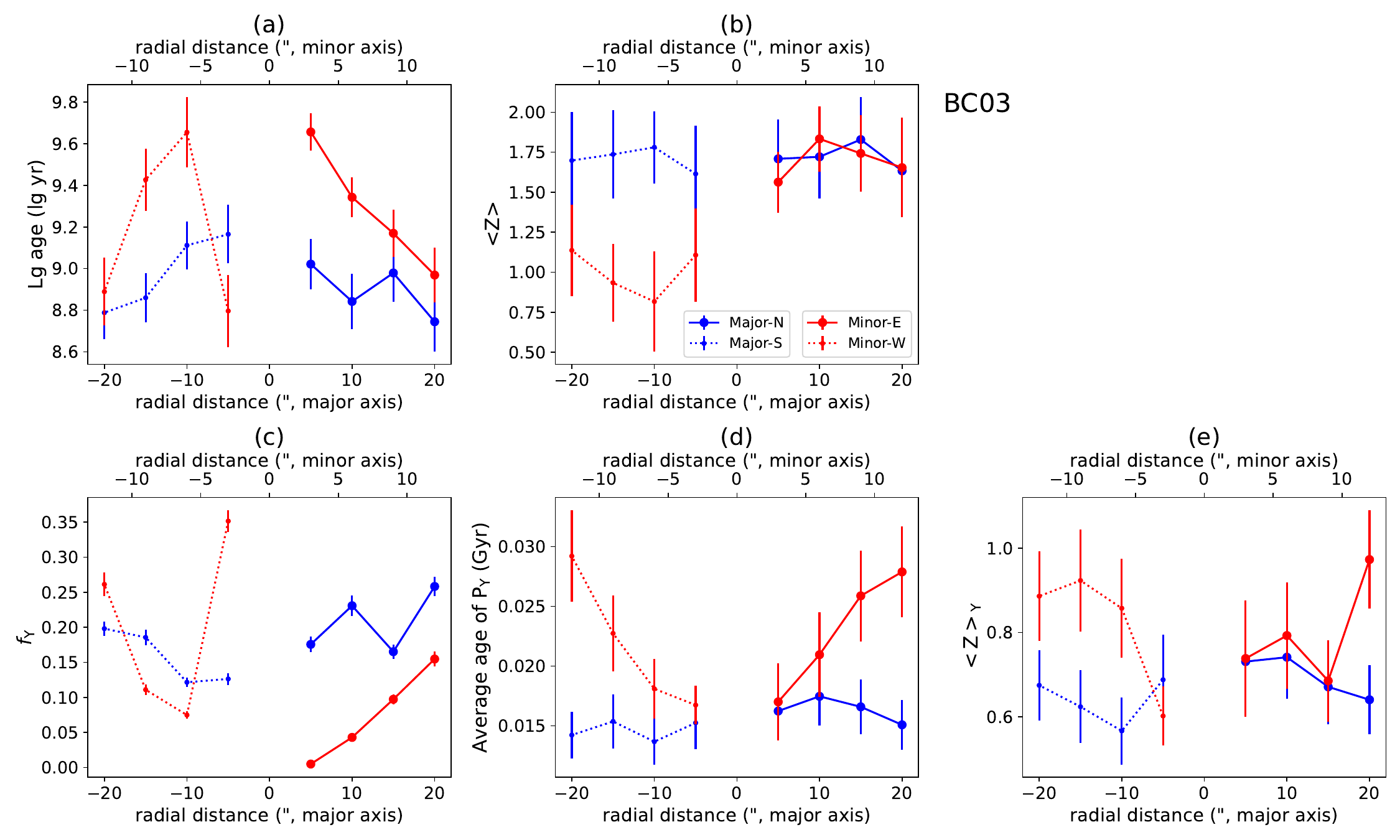}
\caption{Same as Figure~\ref{fig:fsps_rprofile}, but for radial profiles obtained from BC03 synthesis.}
\label{fig:bc03_rprofile}
\end{figure*}

Figure~\ref{fig:fsps_rprofile} shows the radial profiles of the average stellar age, metallicity, $f_{\rm Y}$, and the mean age and metallicity ($<Z>_{\rm Y}$) of the young stellar population $P_{\rm Y}$ obtained from FSPS synthesis.
The errors are estimated from the standard deviation of the resulting stellar population properties obtained from the bootstrapping method (see Sect~\ref{subsec:ste-pop syn}).
Note that the innermost point in the west direction along the minor axis (Minor-W) falls within the dust lane, where $f_{\rm Y}$ is high. 
This point is excluded when analyzing trends along the minor axis.
In Figure~\ref{fig:fsps_rprofile}(a), negative trends of stellar age are observed along both the major and minor axes, consistent with an inside-out star formation scenario.
In Figure~\ref{fig:fsps_rprofile}(b), $<Z>$ shows positive trends in all directions except for the north of the major axis.
Regarding $P_{\rm Y}$, $f_{\rm Y}$ exhibits a positive trend along both the major and the minor axes (Figure~\ref{fig:fsps_rprofile}(c)).
Additionally, excluding the innermost and outmost point of Minor-W, $f_{\rm Y}$ along the major axis is higher than that along the minor axis.
The average age of $P_{\rm Y}$ exhibits a positive trend along both the major and minor axes.
$<Z>_{\rm Y}$ derived from FSPS synthesis exhibits weak positive trends along the minor axis and a positive trend along the major axis in the south direction. 
However, in the north direction, a significant negative trend is observed.

Figure~\ref{fig:bc03_rprofile} presents the radial profiles obtained from BC03 synthesis.
The trends of the stellar age (Figure~\ref{fig:bc03_rprofile}(a)) and $f_{\rm Y}$ (Figure~\ref{fig:bc03_rprofile}(c)) are similar to the results obtained from FSPS synthesis.
$<Z>$ is notably lower along the west direction compared to other directions.
The average age of $P_{\rm Y}$ presents a positive trend along the minor axis, while no significant trend is observed along the major axis (Figure~\ref{fig:bc03_rprofile}(d)).
Along the minor axis, the average age of $P_{\rm Y}$ is significantly higher than that along the major axis (Figure~\ref{fig:bc03_rprofile}(d)), in contrast to the findings of the FSPS synthesis (Figure~\ref{fig:fsps_rprofile}(d)).
When compared to the results in Figure~\ref{fig:fsps_rprofile}(d) obtained from FSPS synthesis, the average ages of $P_{\rm Y}$ derived from BC03 synthesis are lower along the major axis and in the inner regions along the minor axis.
$<Z>_{\rm Y}$ shows a positive trend along the minor axis, while a weak negative trend is observed along the north direction of the major axis (Figure~\ref{fig:bc03_rprofile}(e)).
Along the major axis, $<Z>_{\rm Y}$ is higher compared to that along the minor axis.
Compared to Figure~\ref{fig:fsps_rprofile}(e) obtained from FSPS synthesis, $<Z>_{\rm Y}$ derived from BC03 synthesis is lower along both the major and minor axes, except in the north direction of the major axis.

When comparing the radial profiles derived from BC03 and FSPS synthesis, although there exists some conflicting results, some main findings are generally consistent. 
The trends of the stellar age and $f_{\rm Y}$ obtained from these two syntheses are similar, with $f_{\rm Y}$ derived from FSPS generally higher.
The average age of $P_{\rm Y}$ shows positive trends along the minor axis in both BC03 and FSPS synthesis, although the trend obtained from FSPS is weaker.
Positive trends of $<Z>_{\rm Y}$ along the minor axis and a negative trend along the north direction of the major axis are observed in both syntheses, while the positive trend from FSPS and the negative trend from BC03 are relatively weaker.
These consistent findings about $P_{\rm Y}$ suggest that the young stellar population in the inner regions is younger and metal poorer than that in the outer regions, except for the north direction of the major axis.

\section{Discussion} \label{sec:discussion}

\subsection{Possible AGN feedback on the star formation in NGC 2992} \label{sec:AGN_Fb_ste}

\begin{figure*}[ht!]
\includegraphics[width=1.1\textwidth,trim=30 0 0 30]{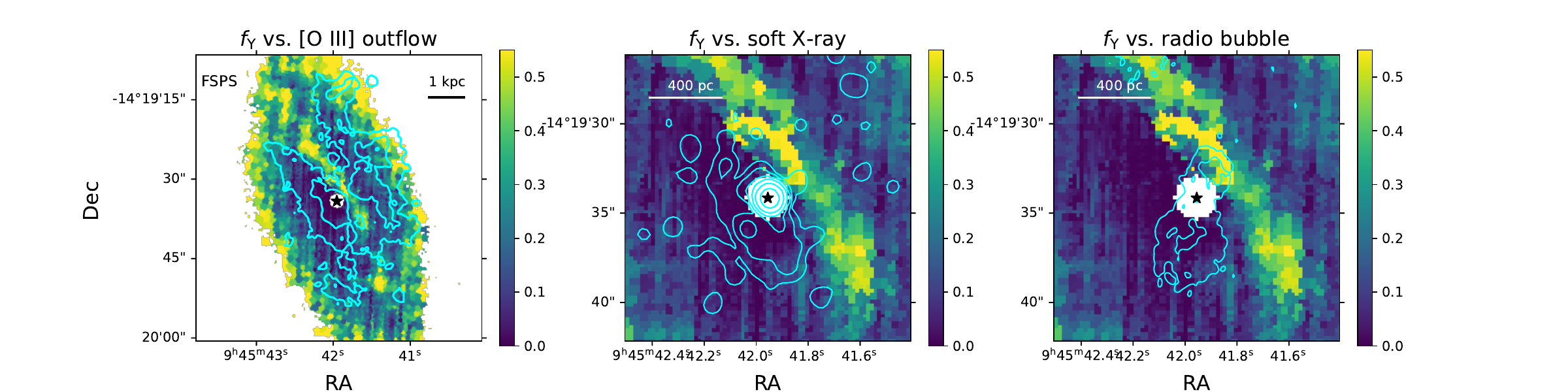}
\caption{Left: $f_{\rm Y}$ map derived from FSPS synthesis (Figure~\ref{fig:fsps_stepop}(c)) superposed with the contours of [O~{\sc{iii}}] outflow component.
The [O~{\sc{iii}}] contours corresponding to 1000 and 5000 $\times 10^{-20}\,\rm erg\,s^{-1}\,cm^{-2}$.
Middle: Central region of $f_{\rm Y}$ map superposed with the contours of soft X-ray emission from \citet{2022ApJ...938..127X}.
The lowest X-ray contour level is $\rm 0.09\, counts/pixel$, and the contour values increase by a factor of 2.9.
Right: Same as the middle panel but superposed with the figure-eight-shaped radio bubble from \cite{1989ApJ...343..659U}.
The radio contour is $\rm 0.0003\,Jy/beam$. 
}
\label{fig:ysp_outflow}
\end{figure*}

In Figure~\ref{fig:ysp_outflow}, we compare the distribution of $f_{\rm Y}$ with the [O~{\sc{iii}}] outflow component, extended soft X-ray emission reported by \citet{2022ApJ...938..127X}, and the nuclear radio bubble documented by \citet{1989ApJ...343..659U}.
\citet{2005ApJ...628..113C} and \citet{2022ApJ...938..127X} found that there exists extended soft X-ray emission around the nucleus. 
\citet{2022ApJ...938..127X} further revealed that the circumnuclear ($r\sim 750\,\rm pc$) soft X-ray emission could be dominated by the hot gas associated with the AGN outflow traced by the nuclear figure-eight-shaped radio bubble.

In the northern region near the nucleus, the highest $f_{\rm Y}$ (median $f_{\rm Y}\sim 0.6$) is observed at the outskirts of the northern radio bubble (right panel of Figure~\ref{fig:ysp_outflow}), indicating an AGN positive feedback scenario.
This finding is similar to the result detected by \citet{2023A&A...678A.127V} in the Teacup galaxy, where a younger stellar population was identified at the periphery of a large ionized gas bubble compared to the rest of the galaxy.
Except for the northern and western regions, the circumnuclear region ($<1$ kpc) exhibits the lowest $f_{\rm Y}$ ($<0.01$) but strong soft X-ray and [O~{\sc{iii}}] outflow component emission.
There is a significant amount of molecular gas in this region \citep{2023A&A...679A..88Z}, but very few young stars have formed.
This result also suggests a possible AGN negative feedback scenario in the circumnuclear region.

\subsubsection{Coupling efficiency} \label{subsec:coup_eff}

Some numerical simulations suggest that the coupling efficiency $\dot{E}_{\rm out}/L_{\rm bol}$ should be larger than $0.5\%$ \citep[e.g.][]{2010MNRAS.401....7H} to exert a noticeable feedback effect on the host galaxy.
The AGN in NGC 2992 is a well-known source with intense variability, exhibiting X-ray 2-10 keV luminosity variations between $3.8\times 10^{41}\rm\,erg\,s^{-1}$ and $1.4\times 10^{43}\rm\,erg\,s^{-1}$ over timescales of decades \citep[e.g.][]{2018MNRAS.478.5638M,2022MNRAS.514.2974M}.
Applying the bolometric correction from \citet{2004MNRAS.351..169M}, we can estimate the bolometric luminosity of $L_{\rm bol} = 2.7\times 10^{42}\rm\,erg\,s^{-1}$--$1.6\times 10^{44}\rm\,erg\,s^{-1}$.

Recently, \citet{2023A&A...679A..88Z} estimated the kinetic power of the ionized gas outflow to be $\dot{E}_{\rm out} = 1.2\times 10^{42}\rm\,erg\, s^{-1}$ from [O~{\sc{iii}}] emission line using the same MUSE data.
By adopting this value, the coupling efficiency can range from $0.75\%$ to $45\%$, depending on the bolometric luminosity.
Even though the lower limit of the coupling efficiency is larger than the threshold suggested by \citet{2010MNRAS.401....7H}.
Hence, the galactic ionized gas outflow in NGC 2992 could impact the host galaxy according to the computed coupling efficiency.

\subsubsection{Timescales} \label{subsec:timescale}

The dynamical timescale of the ionized gas outflow can be estimated by $t_{\rm out}=d/v$, where $d$ is the size of the outflow and $v$ is the average velocity of the outflow.
Assume that $d = 7 \rm\,kpc$ and $v = 200 \rm\,km\,s^{-1}$ \citep[see also][]{2023A&A...679A..88Z}, we can calculate $t_{\rm out} \sim 34 \rm\,Myr$.
This dynamical timescale is close to the timescale of the young stellar population, which is less than $40\rm\,Myr$ according to both BC03 and FSPS synthesis (Figure~\ref{fig:bc03_rprofile} and Figure~\ref{fig:fsps_rprofile}).
\citet{2001A&A...378..787G} identified two [O~{\sc{iii}}] arcs moving at a velocity of $\sim 70 \rm\, km/s$, consistent with the northwest and southeast bubbles observed in the nuclear radio emission.
By assuming that the radio bubbles are expanding at a velocity of $\sim 70 \rm\, km/s$ with a size of $\sim 700 \rm\, pc$, we can estimate the expansion timescale of the radio bubble to be $\sim 10\rm\, Myr$, which aligns closely with the timescale of the young stellar population.
This suggests that the AGN outflow could affect the formation of the young stellar population.

\subsection{Interaction between NGC 2992 and NGC 2993} \label{subsec:galaxy interact}

In simulations, major merger processes can induce gas inflow towards the central regions of galaxies, thereby boosting circumnuclear star formation and triggering nuclear activities \citep[e.g.][]{2010MNRAS.407.1529H}. 
The star formation rate in merging galaxies is expected to increase, particularly in the central region \citep[e.g.][]{2013MNRAS.435.3627E,2021ApJ...909..120S}.
Additionally, in galaxy mergers, the metallicity gradient may be altered as a result of gas inflow \citep[e.g.][]{2019MNRAS.482L..55T}.

\citet{2000AJ....120.1238D} suggested that NGC 2992 and NGC 2993 are observed at an early stage of the interaction, $\sim$100 Myr after the first pericentre passage.
The timescale of interaction is estimated to be 2-3 times longer than that of the young stellar population and ionized outflow (see Sect~\ref{subsec:timescale}). 
However, considering the time required for the gas to respond, the interaction between these galaxies could potentially trigger the formation of the young stellar population and AGN activities in NGC 2992.
\citet{2010A&A...519A..79F} suggested that galaxy interaction could trigger nuclear star formation. 
Additionally, \citet{2021MNRAS.502.3618G} discovered a young, metal-poor stellar population in the nucleus of NGC 2992, proposing that gas inflows resulting from the interaction between NGC 2992 and NGC 2993 stimulate both star formation and AGN activities.

Although the metallicity derived from FSPS and BC03 synthesis are different, the trends of $<Z>_{\rm Y}$ along the north direction of the major axis obtained from both syntheses show negative trends, potentially indicating gas inflow in this region.
This region is situated at the base of the tidal tail connecting NGC 2992 and the tidal dwarf galaxy Arp 245N (Figure~\ref{fig:arp_245}).
Previous studies have indicated that the SFR can be enhanced at the base of tidal tails \citep[e.g.][]{2009AJ....137.4643H,2014AJ....147...60S}.
The $f_{\rm Y}$ in the northern region of the stellar disk is higher than that in the southern region, suggesting that the star formation may be enhanced by gas inflow in the northern region.

\subsection{Contamination from the AGN continuum?}

The correlation between the ionization cones and the distribution of stellar populations raises concerns about potential contamination from the AGN continuum, which could impact the accuracy of the fitting results.
\citet{2021MNRAS.502.3618G} conducted tests on the AGN continuum contamination in the nuclear region of NGC 2992 using GMOS data. 
They utilized a featureless power-law continuum ($F_{\lambda}\propto \lambda^{1.7}$) to account for AGN emission and combined it with stellar population templates to fit the spaxels within the nuclear region ($r<1.1\arcsec$). 
Their findings indicate that the contribution of the AGN continuum to the total light diminishes with increasing radius, becoming negligible beyond $r\sim 1\arcsec$. 
In the nuclear region, where the AGN contribution is significant, the derived stellar populations would appear younger if the featureless continuum is not included in the fitting process \citep[][]{2021MNRAS.502.3618G}. 
Based on their results, the AGN continuum contribution outside the nuclear region can be ignored. 

We also tested the contribution of the AGN continuum in some spaxels by combining a power law component ($F_{\lambda}\propto \lambda^{1.7}$) in the continuum fitting. 
As expected, the power-law component was negligible in these spectra. 
If we assume that the AGN continuum does impact the continuum fitting, it would lead to an underestimation of the mean age of stellar populations in the ionization cones, thereby strengthening the findings of this study.

\section{Conclusions} \label{sec:summary}

In this work, we study the spatially resolved properties of stellar populations of NGC 2992 using the archival VLT/MUSE data. 
Two stellar templates are employed to fit the stellar continuum. 
The extinction in the V band and the stellar kinematics obtained from the continuum fitting are shown.
The spatially resolved maps and radial profiles along the major and minor axes of light-weighted stellar age, light-weighted metallicity $<Z>$, fraction of the young stellar population $f_{\rm Y}$ (age $< 100\,\rm Myr$), and average age and metallicity ($<Z>_{\rm Y}$) of the young stellar population are presented.
The main results and conclusions are summarized as follows:

\begin{enumerate}
    \item Both the results obtained from BC03 and FSPS synthesis indicate the presence of a very young stellar population (age $<40$ Myr) within the dust lane and nearly along the major axis.
    The metallicities derived from the two stellar population syntheses differ, with BC03 synthesis indicating a super-solar metallicity across the stellar disk, while FSPS synthesis reveals a sub-solar metallicity.

    \item Positive trends of the average stellar age and $f_{\rm Y}$ are found along both major and minor axes, indicative of an inside-out star formation scenario.
    Positive trends of $<Z>_{\rm Y}$ are observed in all directions except for the north of the major axis in both BC03 and FSPS synthesis.
    Along the northern direction of the major axis, $<Z>_{\rm Y}$ displays a negative trend, suggesting a gas inflow in this region.

    \item Along the minor axis, where the AGN outflow is prominent, the fraction of the young stellar population ($f_{\rm Y}$) is low.
    In the circumnuclear region, the highest value of $f_{\rm Y}$ is found at the outskirts of the northern radio bubble (Figure~\ref{fig:ysp_outflow}).
    Based on discussions regarding the coupling efficiency and timescales, we propose that the AGN outflow in NGC 2992 might have both negative and positive feedback effects on the formation of young stars especially in the circumnuclear region ($<1$ kpc) (Sec~\ref{sec:AGN_Fb_ste}).

    \item The young stellar population is detected along the major axis of the galaxy, showing spatial alignment with the dust lane and molecular gas disk. 
    This observation could be linked to the interaction between NGC 2992 and NGC 2993, potentially causing gas inflow in the gas disk, consequently triggering the formation of young stars.
\end{enumerate}

In the future, further multi-wavelength observational studies are needed to validate the results of this work.

\section{Acknowledgments}
\begin{acknowledgments}

We thank the anonymous referee for helpful comments that significantly improved the clarity of our work.
We thank Dr. Yulong Gao, Tianwen Cao, Ran Wang, and Siyi Feng for their helpful discussion. 
We acknowledges the National Key R\&D Program of China (Grant No. 2023YFA1607904 and No. 2022YFF0503402).
X.X. acknowledges the China Postdoctoral Science Foundation (No. 2023M741639).
J.W. acknowledges the NSFC grants 12333002, 12033004, and 12221003.
Z.L. acknowledges the National Natural Science Foundation of China (grant 12225302).
Based on observations collected at the European Southern Observatory under ESO program 094.B-0321.

\end{acknowledgments}

%

\vspace{5mm}
\facilities{VLT (MUSE)}


\software{astropy \citep{2013A&A...558A..33A,2018AJ....156..123A}, EsoReflex \citep{2013A&A...559A..96F}, Vorbin \citep[][]{2003MNRAS.342..345C}, PPXF \citep{2023MNRAS.526.3273C}, DS9 \citep{2003ASPC..295..489J},
          }


\bibliography{2992}{}

\bibliographystyle{aasjournal}


\end{CJK*}
\end{document}